\documentclass[%
 reprint,
 amsmath,amssymb,
 aps,
]{revtex4-1}

\usepackage{siunitx}
\usepackage{graphicx}
\usepackage{dcolumn}
\usepackage{bm}

\begin{document}

\preprint{APS/123-QED}

\title{Microresonator solitons for massively parallel coherent \\ optical communications}

\author{Pablo Marin-Palomo\textsuperscript{1,\dag}, Juned N. Kemal\textsuperscript{1,\dag}, Maxim Karpov\textsuperscript{2}, Arne Kordts\textsuperscript{2}, Joerg Pfeifle\textsuperscript{1}, Martin H. P. Pfeiffer\textsuperscript{2}, Philipp Trocha\textsuperscript{1}, Stefan Wolf\textsuperscript{1}, Victor Brasch\textsuperscript{2}, Miles H. Anderson\textsuperscript{2}, Ralf Rosenberger\textsuperscript{1}, Kovendhan Vijayan\textsuperscript{1}, Wolfgang Freude\textsuperscript{1,3}, Tobias J. Kippenberg\textsuperscript{2,*}, Christian Koos\textsuperscript{1,3,*}\\
 }

\affiliation{%
\textsuperscript{\dag}
P.M. and J.N.K contributed equally to this work\\
\textsuperscript{*}
christian.koos@kit.edu 
\textsuperscript{*} 
tobias.kippenberg@epfl.ch\\
\textsuperscript{1} Institute of Photonics and Quantum Electronics (IPQ), Karlsruhe Institute of Technology (KIT), 76131 Karlsruhe, Germany
}%
\affiliation{
\textsuperscript{2} \'{E}cole Polytechnique F\'{e}d\'{e}rale de Lausanne (EPFL), 1015 Lausanne, Switzerland
}%
\affiliation{%
\textsuperscript{3} Institute of Microstructure Technology (IMT), Karlsruhe Institute of Technology (KIT), 76131 Karlsruhe, Germany
}%


\begin{abstract}
Optical solitons are waveforms that preserve their shape while propagating, relying on a balance of dispersion and nonlinearity\cite{Hasegawa1995,Mollenauer1980}. Soliton-based data transmission schemes were investigated in the 1980s, promising to overcome the limitations imposed by dispersion of optical fibers. These approaches, however, were eventually abandoned in favor of wavelength-division multiplexing (WDM) schemes that are easier to implement and offer improved scalability to higher data rates. Here, we show that solitons may experience a comeback in optical communications, this time not as a competitor, but as a key element of massively parallel WDM. Instead of encoding data on the soliton itself, we exploit continuously circulating dissipative Kerr solitons (DKS) in a microresonator\cite{Herr2014,Akhmediev2008}. DKS are generated in an integrated silicon nitride microresonator \cite{Levy2009} by four-photon interactions mediated by Kerr nonlinearity, leading to low-noise, spectrally smooth and broadband optical frequency combs\cite{Brasch2015}. In our experiments, we use two interleaved soliton Kerr combs to trans-mit a data stream of more than 50~Tbit/s on a total of 179 individual optical carriers that span the entire telecommunication C and L bands. Equally important, we demonstrate coherent detection of a WDM data stream by using a pair of microresonator Kerr soliton combs – one as a multi-wavelength light source at the transmitter, and another one as a corresponding local oscillator (LO) at the receiver. This approach exploits the scalability advantages of microresonator soliton comb sources for massively parallel optical communications both at the transmitter and receiver side. Taken together, the results prove the significant potential of these sources to replace arrays of continuous-wave lasers in high-speed communications. In combination with advanced spatial multiplexing schemes\cite{Bozinovic2013, Puttnam2015} and highly integrated silicon photonic circuits\cite{Dai2014}, DKS combs may bring chip-scale petabit/s transceivers into reach.
\end{abstract}

\maketitle



The first observation of solitons in optical fibers\cite{Mollenauer1980} in 1980 was immediately followed by major research efforts to harness such waveforms for long-haul com-munications\cite{Hasegawa1995}. In these schemes, data was encoded on soliton pulses by simple amplitude modulation using on-off-keying (OOK). However, even though the viability of the approach was experimentally demonstrated by transmission over one million kilometres\cite{Nakazawa1991}, the vision of soliton-based communications was ultimately hindered by difficulties in achieving shape-preserving propagation in real transmission systems\cite{Hasegawa1995} and by the fact that nonlinear interactions intrinsically prevent dense packing of soliton pulses in either the time or frequency domain. Moreover, with the advent of wavelength-division multiplexing (WDM), line rates in long-haul communication systems could be increased by rather simple parallel transmission of data streams with lower symbol rates, which are less dispersion sensitive. Consequently, soliton-based communication schemes have moved out of focus over the last two decades.
More recently, frequency combs were demonstrated to hold promise for revolutionizing high-speed optical communications, offering tens or even hundreds of well-defined narrowband optical carriers for massively parallel WDM\cite{Puttnam2015,Ataie2015,Hillerkuss2012}. Unlike carriers derived from a bank of individual laser modules, the tones of a comb are intrinsically equidistant in frequency, thereby eliminating the need for individual wavelength control and for inter-channel guard bands\cite{Puttnam2015,Hillerkuss2012}. In addition, when derived from the same comb source, stochastic frequency variations of optical carriers are strongly correlated, permitting efficient compensation of impairments caused by nonlinearities of the transmission fiber\cite{Temprana2015}. 
For application in optical communications, frequency comb sources must be compact. In recent years, a wide variety of chip-scale comb generators have been demonstrated\cite{Weimann2014,Pfeifle2014}, enabling transmission of WDM data streams with line rates\cite{Kemal2017} of up to 12~Tbit/s. Transmission at higher line rates however, requires more carriers and lower noise levels, and still relies on spectral broadening of narrowband seed combs using dedicated optical fibers\cite{Puttnam2015,Ataie2015,Hillerkuss2012} or nanophotonic waveguides\cite{Hu2016}. In addition, generating uniform combs with a broadband spectral envelope often requires delicate dispersion management schemes, usually in combination with intermediate amplifiers\cite{Ataie2015}. Such schemes are difficult to miniaturize and not readily amenable to chip-scale integration. Moreover, with a few exceptions at comparatively low data rates\cite{Kemal2016}, all advanced comb-based transmission experiments exploit the scalability advantages only at the transmitter, but not at the receiver, where individual continuous-wave (CW) lasers are still used as optical local oscillators (LO) for coherent detection. 

In this paper, we show that dissipative Kerr solitons\cite{Herr2014} (DKS) generated in integrated photonic microresonators via Kerr-nonlinear four-photon interactions, allow for generation of highly stable broadband frequency combs that are perfectly suited to overcome scalability limitations of massively parallel optical transmission both at the transmitter and receiver. Microresonator-based Kerr comb sources\cite{DelHaye2007,Kippenberg2011} intrinsically offer unique advantages such as small footprint, large number of narrow-linewidth optical carriers, and line spacings of tens of GHz, which can be designed to fit established WDM frequency grids. However, while these advantages were recognized, previous transmission experiments\cite{Pfeifle2014} were limited to aggregate line rates of 1.44~Tbit/s due to strong irregularities of the optical spectrum associated with the specific Kerr comb states.

These limitations can be overcome by using DKS combs generated in a CW-driven microresonator. The technique exploits four-wave mixing, a process that converts two pump photons into a pair of new photons. In a microresonator, the strength of this interaction is given by the single-photon Kerr frequency shift of $g = \frac{\hbar\omega_{0}^{2}cn_2}{n^2V_{\text{eff}}}$, where $n_2 (n)$ is the nonlinear (linear) refractive index, $c$ the speed of light in vacuum, $\hbar$ the reduced Planck constant, $V_\text{eff}$ the nonlinear optical mode volume and $\omega_0$ the angular frequency of the pump mode. For pump powers that cause the cavity Kerr frequency shift to exceed the linewidth of the resonance, parametric oscillations lead to the emergence of Kerr frequency combs (modulation instability, MI Kerr combs). Bright DKS can be formed in this process when the microresonator has a locally anomalous group velocity dispersion (i.e. a Taylor expansion for cavity modes $\omega_\mu = \omega_0 + \sum_{i = 1}^{\infty} \frac{D_i}{i!}\mu^i$ has $D_2 > 0$) and a CW laser excitation (frequency $\omega_L$) that is red detuned from the cavity resonance (i.e. $\delta\omega = \omega_0 - \omega_L > 0$). Mathematically, DKS states appear as specific solutions of the Lugiato-Lefever equation\cite{Lugiato1987} and consist of an integer number of discrete secant-hyperbolic-shaped pulses on a CW background circulating in the cavity\cite{Herr2014}. Stability of such states relies on the double balance of dispersion and Kerr nonlinearity as well as of nonlinear parametric gain and cavity loss. Theoretically predicted in Ref. \cite{Akhmediev2008} and first reported to spontaneously form in microresonators made from crystalline MgF$_2$, DKS have been observed in different types of material systems including silica-on-silicon\cite{Yi2015} and silicon nitride\cite{Brasch2015} ($\text{Si}_3\text{N}_4$) as well as silicon\cite{Yu2016}. Of particular interest are single-soliton states, which consist of only one ultra-short pulse circulating around the cavity, leading to a broadband comb spectrum with a smooth and numerically predictable\cite{Herr2014} envelope given by $P(\mu) \approx \frac{\kappa_{\text{ex}}D_2\hbar\omega_0}{4g}\text{sech}^2(\frac{\pi\mu}{2}\sqrt{\frac{D_2}{2\delta\omega}})$, where $\kappa_\text{ex}$ represents the external coupling rate to the bus waveguide (see SI for more details). DKS have already been used in applications such as self-referencing of optical frequency combs\cite{Jost2015,Brasch2017}, low noise microwave generation\cite{Liang2015} and dual soliton comb spectroscopy\cite{Yi2015,Suh2016}.

Our demonstration comprises a series of three experiments that exploit the extraordinarily smooth and broadband spectral envelope and the inherently low phase noise of DKS combs for massively parallel coherent communications. In a first experiment, we transmit data on 94 carriers that span the entire telecommunication C and L bands with a line spacing of $\sim$~100 GHz. Using 16-state quadrature amplitude modulation (16QAM) to encode data on each of the lines, we achieve an aggregate line rate of 30.1~Tbit/s. In a second experiment, we double the number of carriers by interleaving two DKS combs. This gives a total of 179 carriers and an aggregate line rate of 55.0~Tbit/s transmitted over a distance of 75~km – the highest data rate achieved to date with a chip-scale frequency comb source. In a third experiment, we demonstrate coherent detection using a DKS comb as a multi-wavelength LO. We use 93 carriers to transmit and receive an aggregated line rate of 37.2~Tbit/s. In these experiments, the LO comb is coarsely synchronized to the transmitter comb, and digital signal processing is used to account for remaining frequency differences. The results prove the tremendous potential of Kerr soliton combs, both as multi-wavelength optical sources and as LO for massively parallel WDM transmission. 

Our work relies on integrated $\text{Si}_3\text{N}_4$ microresonators\cite{Herr2014} for generation of DKS frequency combs, cf. Fig.~1a. The $\text{Si}_3\text{N}_4$ platform is chosen due to its low optical losses and its compatibility with large-scale silicon-based processing\cite{Levy2009}. The microresonators feature a waveguide height of 800~nm to achieve anomalous GVD and are fabricated using the photonic Damascene process\cite{Pfeiffer2016}. Neighboring resonances are spaced by 100~GHz while featuring intrinsic \textit{Q}-factors exceeding $10^6$. DKS combs are obtained by sweeping of the pump laser through the resonance from a blue-detuned wavelength to a predefined red-detuned wavelength\cite{Herr2014,Brasch2015}. This leads initially to the generation of MI Kerr combs followed by DKS states once the resonance is crossed,  cf. Fig.~1. Importantly, once a multiple-soliton comb state is generated, the transition to a single-soliton state can be accomplished in a reliable and deterministic manner\cite{Guo2017} by adjusting the laser frequency cf. Fig.~1a and 1b and Methods. The measured power spectrum of the DKS comb state is shown in Fig.~1c, which exhibits a 3~dB spectral bandwidth of $\sim$~6~THz. The soliton comb states are remarkably stable for many hours in a laboratory environment, which is key to the transmission experiments presented in this work.

\begin{figure*}
\includegraphics[trim={0.2cm 0 0 0},scale=0.9,clip]{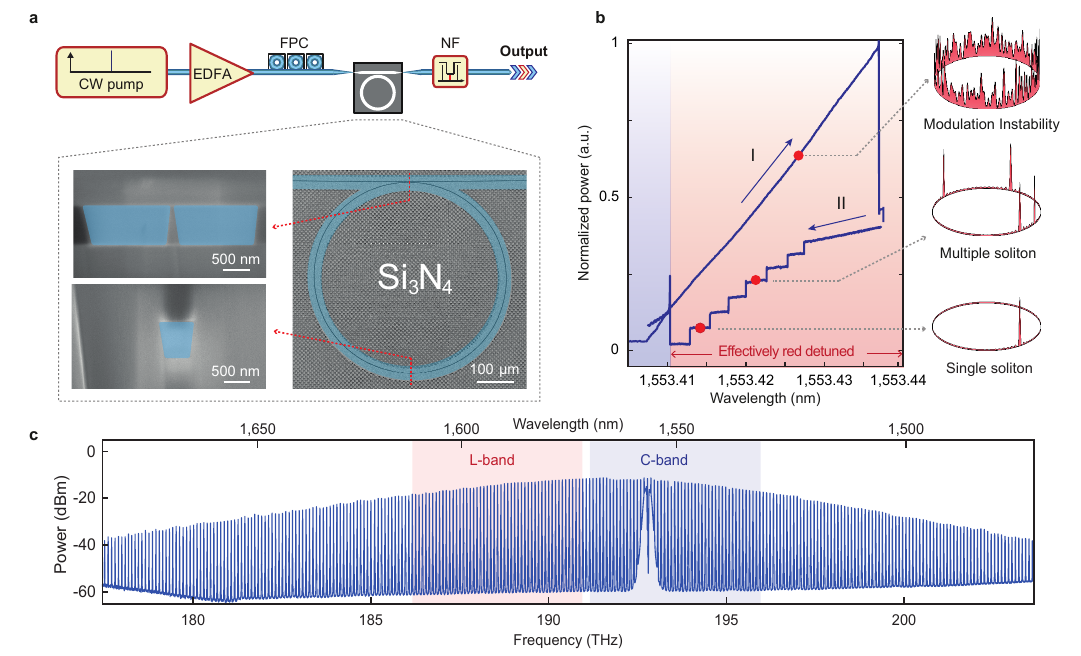}
\caption{\label{fig:1}\textbf{Broadband Kerr comb generation using dissipative Kerr solitons in high-\textit{Q} silicon nitride microresonators.} \textbf{a}, Principle of soliton frequency comb generation: The integrated photonic microresonator is pumped by a tunable CW laser amplified by an erbium-doped fiber amplifier (EDFA). Lensed fibers are used to couple light to the chip (FPC – fiber polarization controller). After the microresonator, a notch filter (NF) suppresses the remaining pump light. The insets show scanning electron microscopy images of a $\text{Si}_3\text{N}_4$ microresonator with a radius of \SI{240}{\micro\metre}. Right inset: Top view. The checker-board pattern results from the photonic Damascene fabrication process\cite{Pfeiffer2016}, see Methods. Left insets: Cross sections of the resonator waveguide ($0.8\times1.65~\SI{}{\micro\metre}^2$) at the coupling point (top) and at the tapered section (bottom, dimensions $0.8\times0.6~\SI{}{\micro\metre}^2$). The tapered section is used for suppressing higher order modes families\cite{Kordts2015}, while preserving a high optical quality factor ($Q\sim10^{6}$), see Methods. \textbf{b}, Pump tuning method for soliton generation in optical microresonator, showing the evolution of the generated comb power versus pump laser wavelength: (I) The pump laser is tuned over the cavity resonance from the blue-detuned regime,  where high-noise modulation instability (MI) combs are observed, to the red-detuned regime, where cavity bistability allows for the formation of soliton states (here a multiple soliton state); (II) Upon having generated a multiple soliton state, the pump laser is tuned backwards to reduce the initial number of solitons down to a single one. The insets on the right schematically show the corresponding intracavity waveforms in different states (MI, multiple and single soliton state); \textbf{c}, Measured spectrum of a single-soliton frequency comb after suppression of residual pump light. The frequency comb features a smooth spectral envelope with a 3~dB bandwidth of 6~THz comprising hundreds of optical carriers extending beyond the telecommunication C and L bands (blue and red, respectively).}
\end{figure*}

The general concept of massively parallel data transmission using a frequency comb as a light source is depicted in Fig.~2a. For emulating massively parallel WDM transmission in our laboratory, we rely on a simplified scheme using only two independent data streams on neighboring channels along with an emulation of polarization division multiplexing, see Section 2 of the Supplementary Information (SI) for details. We use 16QAM at a symbol rate of 40~GBd along with band-limited Nyquist pulses that feature approximately rectangular power spectra, Fig.~2b. At the receiver, each channel is individually characterized using a CW laser as LO and an optical modulation analyzer, which extracts signal quality parameters such as the error-vector magnitude (EVM) or the bit-error ratio (BER). The BER of the transmission experiment are depicted in Fig.~2e with different BER thresholds indicated as horizontal dashed lines. Out of the 101 carriers derived from the comb in the C and L band, a total of 94 channels were used for data trasmission. This leads to a total line rate of 30.1~Tbit/s, see Methods for details on data rate calcuations. The transmission capacity is restricted by the fact that the line spacing of $\sim$~100~GHz significantly exceeds the signal bandwidth of $\sim$~40~GHz, leading to unused frequency bands between neighbouring channels, see Fig.~2b, and hence to a rather low spectral efficiency (SE) of 2.8~bit/s/Hz.

\begin{figure*}
\includegraphics[trim={0.1cm 0.2cm 0.1cm 0.2cm},scale=0.9,clip]{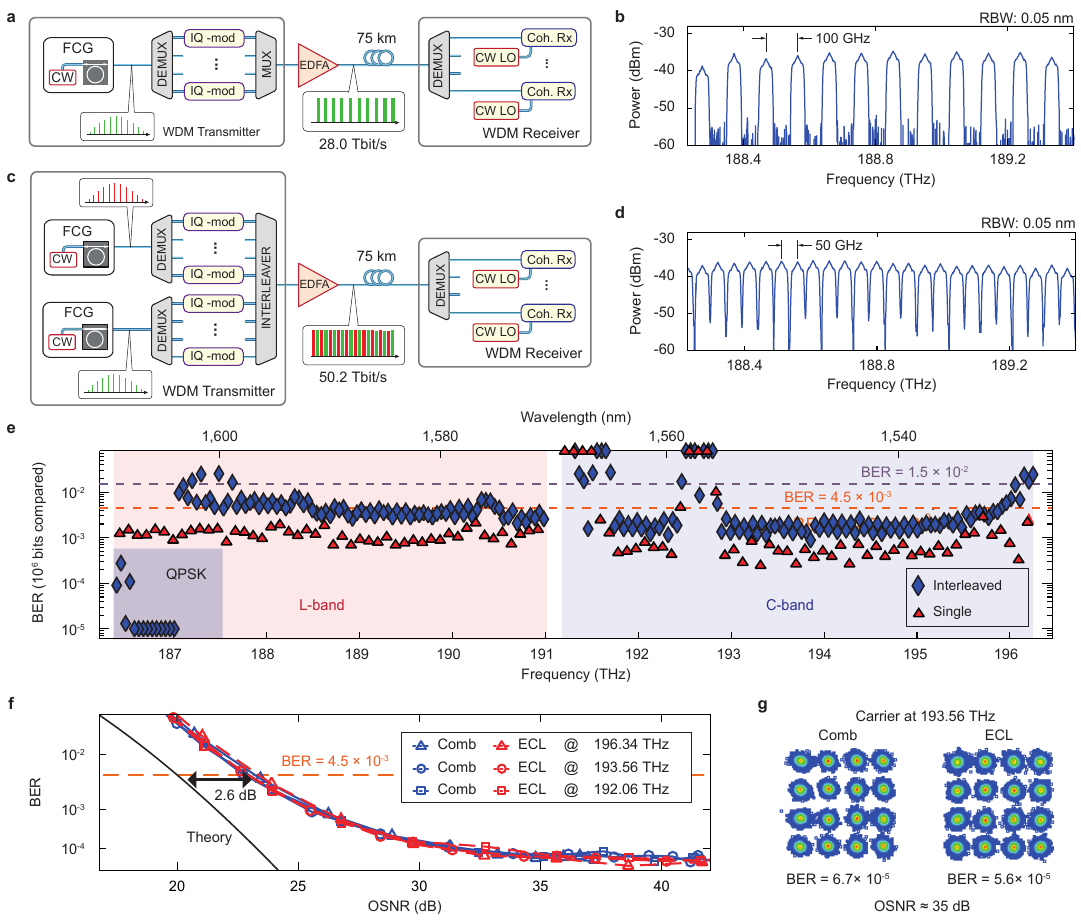}
\caption{\label{fig:2}\textbf{Data transmission using microresonator soliton frequency comb generators as optical sources for massively parallel WDM}. \textbf{a}, Principle of data transmission using a single DKS comb generator as optical source at the transmitter. A demultiplexer (DEMUX) separates the comb lines and routes them to individual dual-polarization in-phase/quadrature (IQ) modulators, which encode independent data on each polarization. The data channels are then recombined into a single-mode fiber using a multiplexer (MUX) and boosted by an erbium-doped fiber amplifier (EDFA) before being transmitted. At the receiver, the wavelength channels are separated by a second DEMUX and detected using digital coherent receivers (Coh. Rx) along with individual CW lasers as local oscillators (CW LO). In our laboratory experiment, we emulate WDM transmission by independent modulation of even and odd carriers using two IQ modulators, see Section~2 of the SI for more details. We use 16-state quadrature amplitude modulation (16QAM) at a symbol rate of 40~GBd per channel. \textbf{b}, Section of the optical spectrum of the WDM data stream. Nyquist pulse shaping leads to approximately 40~GHz wide rectangular power spectra for each of the carriers, which are spaced by $\sim$100~GHz. \textbf{c}, Principle of data transmission using interleaved DKS combs. At the transmitter, two combs of practically identical line spacing are interleaved. The resulting comb features a carrier spacing of $\sim$~50~GHz, which enables dense spectral packing of WDM channels and hence high spectral efficiency. At the receiver, this scheme still relies on individual CW lasers as LO for coherent detection, see Section~2 of SI for details. \textbf{d}, Section of the optical spectrum of the WDM data stream. \textbf{e}, Measured bit-error ratios (BER) of the transmitted channels for the single-comb and the interleaved-comb experiment, along with the BER thresholds\cite{Chang2010} for error-free propagation when applying forward error correction schemes with 7~\% overhead ($4.5\times10^{-3}$, dashed orange line) and 20~\% overhead ($1.5\times10^{-2}$, dashed blue line), see Methods for details. For the interleaved-comb experiment, the outer 14 lines at the low-frequency edge of the L band were modulated with quadrature phase-shift keying (QPSK) signals rather than 16QAM due to the low OSNR of these carriers. 
\textbf{f}, Measured BER vs. optical signal-to-noise ratio OSNR of three different channels derived from a DKS frequency comb (blue) and a high-quality ECL (red), all with 16QAM signalling at 40~GBd. A total of $10^6$ bits were compared. The comb lines do not show any additional penalty in comparison the ECL tones. Similar results were obtained at other symbol rates such as 28~GBd, 32~GBd, or 42.8~GBd. \textbf{g}, Constellation diagrams obtained for an ECL and DKS comb tone at 193.56~THz.}
\end{figure*}

These restrictions can be overcome by using interleaved frequency combs, see Fig.~2c. The scheme relies on a pair of DKS combs with practically identical line spacing, but shifted with respect to each other by half the line spacing by thermal tuning. At the receiver, this scheme still relies on individual CW lasers as LO for coherent detection. The interleaved comb features a line spacing of $\sim$~50~GHz, which allows for dense packing of 40~GBd data channels in the spectrum, see Fig.~2d. The BER results of the transmission experiment are depicted in Fig.~2e with different thresholds indicated as horizontal dashed lines. For a given forward-error correction (FEC) scheme, these thresholds define the maximum BER of the raw data channel that can still be corrected to a BER level below $10^{-15}$, which is considered error-free\cite{Chang2010}, see Methods. We find a total of 204 tones in the C and L band, out of which 179 carriers could be used for data transmission. The remaining channels were not usable due to technical limitations, see Section~2 of the SI. The transmission performance is slightly worse than in the single-comb experiment since twice the number of carriers had to be amplified by the same EDFA, which were operated at their saturation output power such that the power per data channel reduced accordingly. Nevertheless, data was successfully transmitted over 75~km of standard single-mode fiber at a symbol rate of 40~GBd using a combination of 16QAM and QPSK. The total line rate amounts to 55.0~Tbit/s, and the net data rate is 50.2~Tbit/s. This is the highest data rate so far achieved with a chip-scale frequency comb source, and compares very well to the highest capacity of 102.3~Tbit/s ever transmitted through a single-mode fiber core\cite{Sano2012} using more than 200 discrete DFB lasers as optical sources at the transmitter. In addition, we achieve an unprecedented spectral efficiency of 5.2~bit/s/Hz, owing to the densely packed spectrum, Fig.~2d. In the experiments, limited saturation output power of the employed EDFA is the main constraint of signal quality and BER, see Sections~2 and 5.3 of the SI for details. The presented data rates are hence not limited by the DKS comb source, but by the components of the current transmission setup, leaving room for further improvement. 

To confirm the outstanding potential of DKS combs for data transmission, we compare the transmission performance of a single comb line to that of a reference carrier derived from a high-quality benchtop-type external-cavity laser featuring an optical linewidth of approximately 10~kHz, an optical output power of 15~dBm, and an optical carrier-to-noise power ratio (OCNR) in excess of 60~dB. As a metric for the comparison, we use the OSNR penalty at a BER of $4.5\times10^{-3}$, which corresponds to the threshold for FEC with 7~\% overhead\cite{Chang2010}. The results for 40~GBd 16QAM tansmission are shown in Fig.~2f for three different comb lines and for ECL reference transmission experiments at the corresponding comb line frequencies. The OSNR\textsubscript{ref} values are defined for a reference bandwidth of 0.1~nm, see Section~3 of the SI. As shown, no additional OSNR penalty is observed for the frequency comb when compared with the high-quality ECL: For both sources, we observe an OSNR penalty of 2.6~dB with respect to the theoretically required OSNR (black line) for a BER of $4.5\times10^{-3}$. DKS-based light sources can hence dramatically improve scalability of WDM systems without impairing the signal quality under realistic transmission conditions. The error floor in Fig.~2f is attributed to transmitter nonlinearities and electronic receiver noise in our setup. Figure~2g shows the measured constellation diagrams for the ECL and the comb line at 193.56~THz, both taken at the same OSNR of 35~dB. 

To demonstrate the potential of DKS frequency combs as multi-wavelength LO at the receiver, we perform a third experiment shown schematically in Fig.~3a. At the transmitter, a first DKS comb generator with an FSR of $\sim$~100~GHz serves as an optical source. At the receiver, a second DKS comb source having approximately the same FSR is used to generate the corresponding LO tones, each featuring an optical linewidth below 100~kHz. Figures 3b and 3c show a section of the transmitted data spectrum along with the corresponding section of the LO comb. As a reference, the same experiment was repeated using a high-quality ECL with a 10~kHz linewidth as an LO for channel-by-channel demodulation. Overall, an aggregate data rate of 34.6~Tbit/s is obtained. The resulting BER values of all 99 channels for both methods are shown in Fig.~3d. Some of the channels showed signal impairments due to limitations of the available equipment, see Methods. However, we cannot observe any considerable penalty that could be systematically attributed to using the DKS comb as an LO.

\begin{figure*}
\includegraphics[trim={0.2cm 0 0 0},scale=0.9,clip]{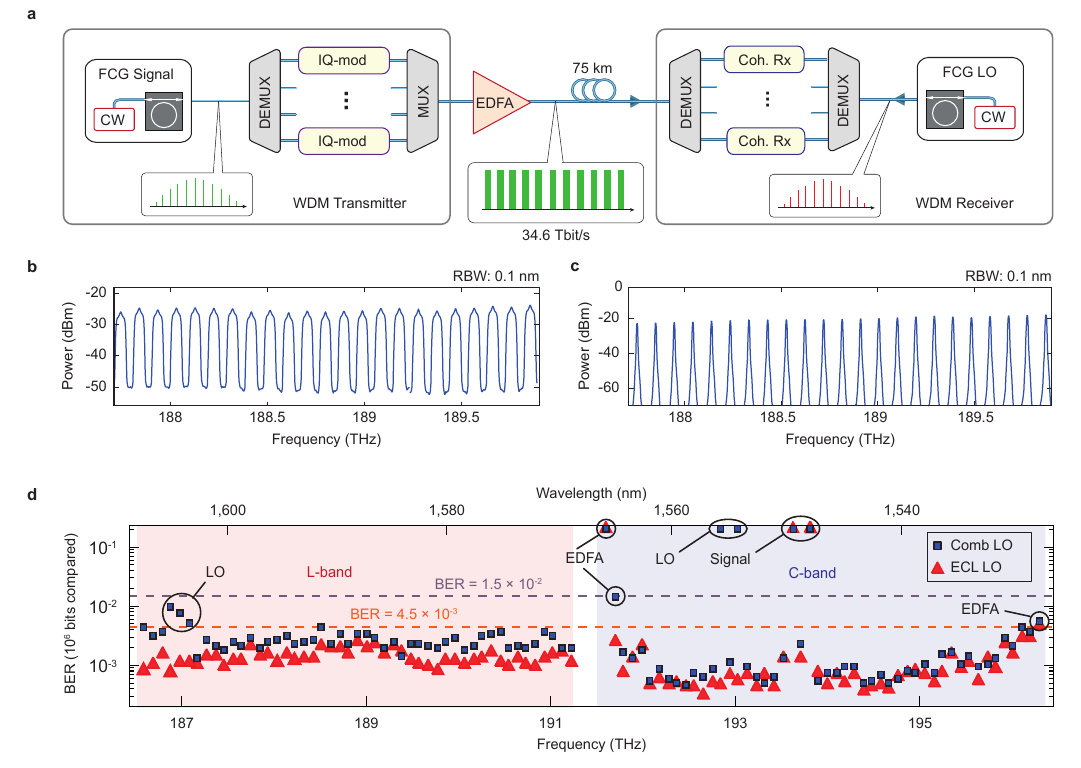}
\caption{\label{fig:3} \textbf{Coherent data transmission using DKS frequency combs both at the transmitter and at the receiver side.}  \textbf{a}, Massively parallel WDM data transmission schematic using DKS frequency combs both as multi-wavelength source at the transmitter and as multi-wavelength local oscillator (LO) at the receiver. We use 16QAM at a symbol rate of 50~GBd per channel. In contrast to Fig.~2a, a single optical source provides all required LO for coherent detection. An extra DEMUX is required to route each LO tone to the respective coherent receiver (Coh. Rx). \textbf{b}, Section of the spectrum of the transmitted channels. \textbf{c}, Corresponding section of the spectrum of the DKS frequency comb used as multi-wavelength LO for coherent detection. The comparatively large width of the spectral lines is caused by the resolution bandwidth (RBW) of the spectrometer (RBW: 0.1~nm). \textbf{d}, Measured BER for each data channel. Blue squares show the results obtained when using a DKS comb as multi-wavelength LO, and red triangles correspond to a reference measurement using a high-quality ECL as LO. Dashed lines mark the BER thresholds of $4.5\times10^{-3}$ ($1.5\times10^{-2}$) for hard-decision (soft-decision) FEC with 7~\% (20~\%) overhead. Black circles show the channels with BER above the threshold for 7~\% FEC and specify the reasons for low signal quality: low OCNR of the carriers from the LO comb (\textquotedblleft LO\textquotedblright) and the signal comb (\textquotedblleft Signal\textquotedblright), as well as bandwidth limitations of the C-band EDFA (\textquotedblleft EDFA\textquotedblright). }
\end{figure*}

While frequency combs offer fundamental technical advantages compared to discrete lasers, they can also contribute to reduce the power consumption of the transmission system. In this context, the power conversion efficiency of the DKS comb generator is an important metric, defined as the ratio between the power of the pump and that of the generated comb lines. The power conversion efficiency of our current comb sources is limited to rather small values between 0.1~\% and 0.6~\% due to the fundamental principle that bright soliton generation only occurs with the pump laser being far detuned from the optical resonance. Still, the overall power consumption can already now compete with massively parallel arrays of commercially available integrated tunable laser assemblies (ITLA), see Methods and Section~5 of the SI for details. The analysis also shows that improvement of more than one order of magnitude compared to ITLA is possible by optimizing the microresonator dispersion, by making use of tailored amplifiers with optimum efficiency, and by increasing the power conversion efficiency via recently demonstrated high-\textit{Q} $\text{Si}_3\text{N}_4$ microresonators with \textit{Q}-factors of $6\times10^6$.  

In summary, we have demonstrated the potential of chip-scale DKS frequency combs for massively parallel WDM at data rates of tens of terabit/s. We use them both as multi-wavelength source at the transmitter and as LO at the receiver, and we show in both cases that that there is no systematic penalty compared to using high-quality ECL. While our experiments achieve the highest data rate with chip-scale frequency comb sources to date, there is still room for increasing the transmission capacity by optimizing the transmission system or by using the adjacent S and U bands for telecommunications in the near infrared. For long transmission distances, comb-based transmission schemes might allow for compensation of nonlinear impairments and hence lead to an improved signal quality compared to conventional WDM schemes\cite{Temprana2015}. The results prove the tremendous potential of DKS comb generators for high-speed data transmission, both in petabit/s intra-datacenter networks\cite{Kachris2012} and in inter-datacenter connections. 


\section*{METHODS}
\subsection*{Fabrication of high-\textit{Q} $\text{Si}_3\text{N}_4$ microresonator}
We use $\text{Si}_3\text{N}_4$ microresonators for generation of DKS frequency combs. These devices lend themselves to co-integration with other photonic devices, using either monolithic approaches on silicon\cite{Dai2014} or indium phosphide (InP)\cite{Nagarajan2010}, hybrid InP-on-silicon approaches\cite{Liang2010}, or multi-chip concepts\cite{Pfeifle2014,Lindenmann2012}. A particularly attractive option is to co-integrate DKS comb generators with advanced multiplexer and demultiplexer circuits\cite{Dai2014} and with highly power-efficient IQ modulators\cite{Koeber2015,Koos2016} to realize chip-scale transceivers that can handle tens of terabit/s of data traffic, as already envisioned in Ref.~\cite{Pfeifle2014}. For soliton formation, anomalous group velocity dispersion (GVD) is required, necessitating $\text{Si}_3\text{N}_4$ waveguide heights of approximately 800~nm for a width of 1.65~$\mu$m. These dimensions are achieved by using the recently developed photonic Damascence process\cite{Pfeiffer2016}, along with a waveguide filtering section to achieve a smooth dispersion profile\cite{Kordts2015}. A mode-filtering section was incorporated into the microrings in order to suppress higher-order modes\cite{Kordts2015}, see Inset of Fig.~1a. This allows to minimize the number of avoided mode crossings and facilitates soliton comb generation. 
Fabrication reproducibility was investigated, leading to a yield of chips enabling DKS generation per batch of approximately 40~\%. For resonators taken from the same wafer, we measured an average line spacing of 95.75~GHz with a standard deviation of approximately 70~MHz. We attribute such differences to variations of both the waveguide width and the thickness of the $\text{Si}_3\text{N}_4$ layer between distant micro-resonators within the same wafer. These variations are approximately 20~nm and may be reduced by increasing the uniformity of the fabrication process over the entire wafer. In particular, we expect that the use of highly developed large-scale fabrication equipment such as 193~nm deep-UV lithography and thin-film tools with improved uniformity will overcome these shortcomings in device reproducibility, as already demonstrated in the context of silicon photonics\cite{Selvaraja2009}. Note that the line spacing difference between frequency combs can be compensate through thermal tuning, which allows us to adapt the microresonators FSR over a tuning range of more than 40~MHz, with a precision of approximately 200~kHz. The same technique can be applied to precisely match the line spacing to established ITU grids. 
In addition, we measured the differences in power conversion efficiency of the DKS comb generators used in our experiment, leading to values that range from 0.1~\% to 0.6~\%. These differences in power conversion efficiency are attributed to variations of the quality factors of the pumped resonances, which may be influenced by interactions of the fundamental mode with higher-order modes. Even though mode-filtering sections are used to suppress the propagation of higher order modes within the waveguide, we can still observe slight variations of the spectral envelope with respect to the theoretical sech\textsuperscript{2} envelope shape. This fact hints to the presence of avoided mode crossings which are characteristic of multimode waveguides. We expect that these limitations can be overcome by optimized device design. 
For DKS comb generators with improver power efficiency, high-\textit{Q} factors are of great importance. Recently, $\text{Si}_3\text{N}_4$ resonators have been demonstrated\cite{Xuan2016,Pfeiffer2017} to reach \textit{Q}-factors of more than $6\times10^6$ both in devices with normal\cite{Xuan2016} and anomalous\cite{Pfeiffer2017} GVD. Along with tailored optical amplifiers, such devices allow to reduce the electrical power consumption by more than an order of magnitude compared to state-of-the-art ITLA laser arrays, see Section~5.1 of the SI.

\subsection*{Soliton comb generation}
The DKS combs are generated by pumping the microresonators with an ECL and a subsequent EDFA, which is operated at an output power of approximately 35~dBm, see Section~1 and Fig.~S1 of the SI for a more detailed description of the comb generation setup. A high-power band-pass filter with a 3~dB bandwidth of 0.8~nm is used to suppress the ASE noise from the optical amplifier. The soliton state is excited by well-controlled wavelength tuning\cite{Herr2014} of the pump ECL from low to high wavelengths across the resonance at a rate of approximately 100~pm/s. Once a multiple soliton state is obtained, the transition to a single-soliton state is accomplished by fine-tuning of the pump laser towards lower wavelengths\cite{Guo2017}. This slow sweep is performed at a rate of approximately 1~pm/s. Light is coupled into and out of the on-chip $\text{Si}_3\text{N}_4$ waveguides by means of lensed fibers, featuring spot sizes of \SI{3.5}{\micro\metre} and coupling losses of 1.4~dB per facet. The power coupled to the chip was approximately 32~dBm. The frequency comb used in the single-comb transmission experiment exhibits a line spacing of 95.80~GHz and a 3~dB bandwidth of more than 6~THz. The optical linewidth of individual comb carriers is limited by the optical linewidth of our tunable pump lasers (TLB-6700, New Focus and TSL-220, Santec), which is below 100 kHz. This value is well below that of telecommunication-grade DFB lasers\cite{Puttnam2014} and hence perfectly suited for coherent communications\cite{Pfau2009}. No additional linewidth broadening relative to the pump is measured, i.e., the phase noise of the comb lines seems to be entirely dominated by the pump. Note, in addition, that an alternative approach for DKS generation has recently been demonstrated\cite{Joshi2016}  where the detuning of the pump laser with respect to the resonance is adjusted by thermally shifting the resonance by means of integrated heaters rather than by tuning the wavelength of the pump laser. This technique allows to replace the tunable laser by much more stable CW pump lasers with sub-kHz linewidths. At the output of the microresonator, a tunable fiber Bragg grating (FBG) acts as a notch filter to suppress the residual pump light to a power level that matches the other comb carriers. After the FBG, the measured optical power of the entire comb spectrum, see Fig.~1c, amounts to 4~dBm. For the experiments using interleaved transmitter (Tx) frequency combs or a separate receiver (Rx) LO comb, a second DKS comb generator with similar performance is used. The frequency comb from the second device for the interleaved Tx combs (for the Rx LO) features a slightly different line spacing of 95.82 GHz (95.70 GHz) and optical power of 0~dBm (8~dBm). For the transmission experiments, an EDFA is used to amplify the combs to an approximate power of 5~dBm per line prior to modulation. The carriers next to the pumped resonance are superimposed by strong amplified stimulated emission (ASE) noise that originates from the optical amplifier. In future implementations, ASE noise can be avoided by extracting the comb light from the microresonator using a drop-port geometry\cite{Wang2016}. This would avoid direct transmission of broadband ASE noise through the device and, in addition, would render the notch filter for pump light suppression superfluous.

\subsection*{Dissipative Kerr soliton comb tuning and interleaving}
Precise interleaving of the frequency combs in the second transmission experiment is achieved by adjusting the temperature of each microresonator, which changes the refractive index and thereby shifts the resonance frequencies while leaving the FSR essentially unchanged\cite{Carmon2004}. A detailed sketch of the experimental setup is given in Section~1 of the SI.  The resonance frequencies of the cavity can be tuned at a rate of approximately $-2.5$~GHz/K with an accuracy of approximately 200~MHz, limited by the resolution of the heater of approximately 0.1~K. In addition, as a consequence of intra-pulse Raman scattering\cite{Karpov2016b}, the center frequency of the comb can also be tuned by slowly changing the pump frequency during operation at a constant external temperature. The associated tuning range is limited to approximately $\pm$500 MHz before the comb state is lost; the tuning resolution is given by the pump laser and amounts to approximately 10~MHz for our devices (TLB-6700, New Focus; TSL-220, Santec). These tuning procedures are used for precise interleaving of DKS combs in the second transmission experiment and for synchronizing the LO comb to the Tx comb in the third transmission experiment.  

\subsection*{Data transmission experimets}
For the data transmission experiments\cite{Pfeifle2015b,Marin2016a,Marin2016b} the single or interleaved frequency comb is amplified to 26.5~dBm by a C/L-band EDFA, before the lines are equalized and dis-interleaved into odd and even carriers to emulate WDM. In the lab experiment, the de-multiplexer (DEMUX) depicted in Fig.~2a is implemented by two programmable filters (Finisar WaveShaper, WS) along with C- and L-band filters, that act as dis-interleavers to separate the combs into two sets of \textquotedblleft even\textquotedblright~and \textquotedblleft odd\textquotedblright~carriers, see Section~2 of the SI for a more detailed description of the experimental setup. After separation, the carriers are routed to individual dual-polarization in-phase/quadrature modulators (IQ-mod), which encode independent data streams on each polarization using both the amplitude and the phase of the optical signal as carriers of information. To this end, we use two optical IQ modulators which are driven with pseudo-random bit sequences of length $2^{11} – 1$ using QPSK or 16QAM signaling and raised-cosine (RC) pulse shaping with a roll-off factor $\beta = 0.1$. The drive signals were generated by arbitrary-waveform generators (AWG). For the transmission experiment using frequency combs as optical source at the Tx, the symbol rate was 40~GBd, and the sampling rate of the AWG was 65~GSa/s (Keysight M8195A). For the experiment with a DKS comb as a multi-wavelength LO, symbol rates of 50~GBd and sampling rates of 92~GSa/s (Keysight M8196A) were used. We refrained from using higher symbol rates since the limited electrical bandwidth of our transmitter and receiver hardware would have led to significantly worse signal quality. In all experiments, polarization division multiplexing (PDM) is emulated by a split-and-combine method, where the data stream of one polarization is delayed by approximately 240~bits with respect to the other to generate uncorrelated data\cite{Hillerkuss2011}. The data channels are then recombined into a 75~km long standard single-mode fiber (SSMF) fiber using a multiplexer (MUX) and boosted by an erbium-doped fiber amplifier (EDFA), before being transmitted. At the receiver, we select each channel individually by a BPF having a 0.6~nm passband, followed by a C-band or an L-band EDFA, and another BPF with a 1.5~nm passband. The signal is received and processed by means of an optical modulation analyzer (OMA, Keysight N4391A), using either a high-quality ECL line or a tone of another DKS comb as local oscillator. In the latter case, we use an optical band-pass filter to extract the tone of interest from the LO comb, see Section~4 of the SI for details. In all experiments, we perform offline processing including filtering, frequency offset compensation, clock recovery, polarization demultiplexing, dispersion compensation, and equalization.

\subsection*{Transmission impairments and data rates}
In our transmission experiments, performance was impaired by specific limitations of the available laboratory equipment, which can be avoided in real-world transmission systems. For the data transmission experiment using a single DKS comb generator as optical source, Fig.~2a, a total of 101 tones were derived from the comb in the C and L band. Out of those, 92 carriers performed better than the BER threshold of $4.5\times10^{-3}$ for widely used second-generation forward-error correction (FEC) with 7~\% overhead\cite{Chang2010}, see Fig.~2e. The pump tone at approximately 192.66~THz and two neighbouring carriers could not be used for data transmission due to strong amplified spontaneous emission (ASE) background from the pump EDFA. Two more directly adjacent channels exceeded the threshold of $4.5\times10^{-3}$, but were still below the BER threshold of $1.5\times10^{-2}$ for soft-decision FEC with 20~\% overhead\cite{Chang2010}. Another four channels at the low-frequency end of the C-band were lost due to a mismatch on the transmission band of the C-band filters used to realize the demultiplexer. This leads to an overall net data rate (line rate) of 30.1~Tbit/s (28.0~Tbit/s). Similarly, for the data transmission experiment using interleaved DKS combs as optical source at the transmiter, Fig.~2b, a total of 126 channels exhibit a BER of less than  $4.5\times10^{-3}$, requiring an FEC overhead of 7~\%, and 39 additional channels showed a BER below  $1.5\times10^{-2}$ which can be corrected by FEC schemes with 20~\% overhead. For the 14 channels at the low-frequency edge of the L band, the modulation format was changed to QPSK since data transmission using 16QAM was inhibited by the low power of these carriers caused by a decrease of amplification of the L-band EDFA in this wavelength range, , see Section~2 of the SI for a more detailed description. This leads to an overall net data rate (line rate) of 50.2~Tbit/s (55.0 ~Tbit/s). For the transmission experiment using DKS frequency combs both at the transmitter and at the receiver side, Fig.~3a, 99 tones were transmitted and tested. Out of those, a total of 89 channels perform better than the BER threshold for hard-decision FEC with 7~\% overhead ( $4.5\times10^{-3}$), and additional four channels are below the BER limit of  $1.5\times10^{-2}$ for soft-decision FEC with 20~\% overhead. For the remaining channels, coherent reception was inhibited by low OSNR. This leads to an overall net data rate (line rate) of 34.6~Tbit/s (37.2~Tbit/s). The black circles in Fig.~3d show the channels with BER above the threshold for 7~\% FEC and specify the reasons for low signal quality: low optical carrier-to-noise ratio (OCNR) of the carriers from the LO comb (LO) and the signal comb (Signal) as well as bandwidth limitations of the C-band EDFA (EDFA). It is further worth noting that field-deployed WDM systems rely on statistically independent data channels rather than on transmitting identical data streams on the \textquotedblleft even\textquotedblright~and the \textquotedblleft odd\textquotedblright~channels. As a consequence, real-world signals will suffer much less from coherent addition of nonlinear interference noise than the signals used in our experiments\cite{Dar2016}. With respect to nonlinear impairments, our experiments hence represent a bad-case scenario with further room for improvement.

\subsection*{Characterization of the OSNR penalty of the frequency comb source}
For comparing the transmission performances of a single comb line to that of a high-quality ECL reference carrier, we measure the OSNR penalty at a BER of     $4.5\times10^{-3}$,which corresponds to the threshold for FEC with 7~\% overhead\cite{Chang2010}. For a given BER, the OSNR penalty is given by the dB-value of the ratio of the actually required OSNR to the OSNR that would be theoretically required in an ideal transmission setup\cite{Schmogrow2012}. 
A detailed description of the associated experimental setup is given in Section~3 of the SI. The carrier under test is selected by a band-pass filter with a 1.3~nm (160~GHz) wide passband. The carrier is then amplified to 24~dBm by an EDFA (EFDA2 in Fig.~S3) and modulated with a PDM-16QAM signal at 40~GBd. Next, an ASE noise source together with two VOA is used to set the OSNR of the channel while keeping its optical power constant. As an ASE generator, we use a second EDFA (EDFA3 in Fig.~S3a). An optical spectrum analyzer (OSA, Ando AQ6317B) is used for measuring the OSNR at the input of the receiver. For each OSNR value, the quality of the channels is determined by measuring the BER using our previously described receiver configuration, see \textquotedblleft Data transmission experiments\textquotedblright~in Methods. At a BER of $4.5\times10^{-3}$, a penalty of 2.6~dB with respect to the theoretical OSNR value is observed, see Fig.~2f, which is a common value for technical implementations of optical 16QAM transmitters\cite{Winzer2010}. For high OSNR, an error floor caused by transmitter nonlinearities and receiver noise is reached. The maximum achievable OSNR of 44~dB at 192.56~THz for transmission with the comb line is dictated by ASE noise of the C/L-band EDFA (EDFA1) right after the FCG, see Fig.~S3. As a reference, the same measurements are repeated using a high quality ECL (Keysight N7714A) to generate the carrier, which leads to essentially the same OSNR penalty for a given BER as the transmission with the comb line. Note that for transmission with the ECL, only one EDFA (EDFA2) is needed to increase the power to 24~dBm before being modulated. As a consequence, a higher maximum OSNR of 58~dB can be achieved with the ECL that with the comb line. Note that for transmission with a single line, the lowest BER reached at 40~GBd falls below $10^{-4}$, as depicted in Fig.~2f. This value, however, is not reached in the WDM transmission experiment with the full comb, Fig.~2a and Fig.~3d. For WDM transmission, a larger number of carriers are amplified by the EDFA in front of the modulator, which, together with the limited output power of the EDFA, leads to a decrease of the optical power per line and hence of the OCNR. In addition, when interleaving two frequency combs, a VOA and a directional coupler are used to interleave the combs and to adapt the power levels. These components introduce additional loss, which needs to be compensated by the subsequent EDFA. Using additional EDFA would therefore increase the quality of the received signal.


\bibliography{Manuscript_Arxiv_2017-04}

\begin{acknowledgments}
This work was supported by the European Research Council (ERC Starting Grant \textquoteleft EnTeraPIC\textquoteright, No. 280145), the EU project BigPipes, the Alfried Krupp von Bohlen und Halbach Foundation, the Karlsruhe School of Optics \& Photonics (KSOP), and the Helmholtz International Research School for Teratronics (HIRST). P.M. is supported by the Erasmus Mundus Doctorate Program Europhotonics (Grant No. 159224-1-2009-1-FR-ERA MUNDUS-EMJD). We further gratefully acknowledge financial support by the Deutsche Forschungsgemeinschaft (DFG) through the Collaborative Research Center  \textquotedblleft Wave Phenomena: Analysis and Numerics\textquotedblright~(CRC 1173), project B3 \textquotedblleft Frequency combs\textquotedblright. $\text{Si}_3\text{N}_4$ devices were fabricated and grown in the Center of MicroNanoTechnology (CMi) at EPFL. EPFL acknowledges support by an ESA PhD fellowship (M.K.) and by the Air Force Office of Scientific Research, Air Force Material Command, USAF, No. FA9550-15-1-0099. M.K. acknowledges funding support from Marie Curie FP7 ITN FACT. M.K. acknowledges funding support from Marie Curie FP7 ITN FACT. In addition, the Swiss National Science Foundation (SNF) is acknowledged, as well as support from the Defense Advanced Research Program Agency through the program QuASAR.\\
\end{acknowledgments}

\section*{AUTHORS CONTRIBUTIONS
}
P.M.-P. and J.N.K contributed equally to this work. P.M.-P., J.N.K, and J.P. built the system for data transmission, supervised by C.K.. Design and fabrication of the $\text{Si}_3\text{N}_4$ microresonators was done by A.K. and M.H.P.P. and supervised by T.J.K.. Samples characterization and selection was done by P.M.-P., M.K., and A.K.. Soliton generation technique was developed by M.K., V.B., A.K. and M.H.P.P. and supervised by T.J.K.. OSNR characterization of the optical source was perform by J.P., P. M.-P. and K.V.. Both data transmission experiments were jointly accomplished by P.M.-P., J.P., P.T., S.W., J.N.K., K.V. and R.R. and supervised by C.K.. Theoretical investigation of the optimization of the power conversion efficiency of DKS sources was done by M.K. and M.H.A.. The project was initiated and supervised by W.F., T.J.K. and C.K.. All authors discussed the data. The manuscript was written by P.M.-P., J.N.K., M.K., T.J.K., and C.K.

\end{document}